\documentclass[sigconf]{acmart}

\usepackage{booktabs} 
\usepackage{enumitem}
\usepackage{graphicx}
\usepackage{epsfig}
\usepackage{amssymb}
\usepackage{amsmath}
\usepackage{xcolor}
\usepackage{xspace}
\usepackage{bbm}
\usepackage{caption}
\usepackage{hyperref}
\usepackage{subcaption}
\usepackage{flushend}
\usepackage{soul}
\usepackage{todonotes}

\setcopyright{rightsretained}

\copyrightyear{2017}
\acmYear{2017}
\setcopyright{rightsretained}
\acmConference{SIGIR 2017 Workshop on Neural Information Retrieval (Neu-IR'17)}{}{August 7--11, 2017, Shinjuku, Tokyo, Japan} \acmPrice{}
\acmDOI{10.475/123_4}
\acmISBN{123-4567-24-567/08/06}

\begin{document}
\title{Towards Semantic Query Segmentation}

\author{Ajinkya Kale, Thrivikrama Taula, Sanjika Hewavitharana, Amit Srivastava}

\affiliation{
  \institution{eBay Inc., 2065 Hamilton Ave, San Jose, CA}}

\email{ {ajkale, ttaula, shewavitharana, amitsrivastava}@ebay.com}

\begin{abstract}
Query Segmentation is one of the critical components for understanding users' search intent in Information Retrieval tasks. It involves grouping tokens in the search query into meaningful phrases which help downstream tasks like search relevance and query understanding. In this paper, we propose a novel approach to segment user queries using distributed query embeddings. Our key contribution is a supervised approach to the segmentation task using low-dimensional feature vectors for queries, getting rid of traditional hand tuned and heuristic NLP features which are quite expensive. 

	We benchmark on a 50,000 human-annotated web search engine query corpus achieving comparable accuracy to state-of-the-art techniques. The advantage of our technique is its fast and does not use external knowledge-base like Wikipedia for score boosting. This helps us generalize our approach to other domains like eCommerce without any fine-tuning. We demonstrate the effectiveness of this method on another 50,000 human-annotated eCommerce query corpus from eBay search logs. Our approach is easy to implement and generalizes well across different search domains proving the power of low-dimensional embeddings in query segmentation task, opening up a new direction of research for this problem.

\end{abstract}

%
%


\keywords{Query Segmentation, Word Embeddings, eCommerce search, Neural Information Retrieval}


\maketitle

\section{Introduction}
  Query Segmentation involves the process of splitting the search query into meaningful continuous segments to assist search precision and retrieval tasks. Search engines fetch high quality and most relevant results when they are able to identify the important phrases in the search query which need to be kept together for quality results. One way to achieve this is if the user is explicit about the phrases by adding quotes around the segments of the search query to indicate phrases. But this is hardly the pattern you see in real world search logs. Users expect the search engine to infer and understand these phrases. This ends up degrading the precision in most cases where the phrase as a whole is important to be kept together during retrieval like movie name, song title, brands etc.  Consider a shopper looking for \textit{long sleeve summer dress}. She is looking for mainly \textit{summer dress} which have \textit{long sleeves}. The underlying search engine needs to know that the query is for a \textit{dress} and specifically for \textit{summer} wear which have \textit{long sleeve} as an additional feature of the dress. The search experience is different if a user searches with quotes around the segments - \textit{"long sleeve" "summer" "dress"} compared to the unquoted query. If the query is treated as a bag of words, the results might end up being less precise. In the \textit{long sleeve summer dress}, the results are far from accurate if we show the user items which match \textit{long summer dress with short sleeves}. Order plays a vital role in query segmentation which is lost in a bag of words model. 

Rest of the paper is organized as follows:  
Section 2 reviews past and recent work on query segmentation followed by the motivation for our novel approach for semantic query segmentation in Section 3. Section 4 will describe our modeling architecture. Section 5 will cover our experiments and results on AOL web query corpus and on annotated eBay eCommerce query corpus. We provide an in-depth analysis and comparison of our techniques with other query segmentation approaches. We close out in Section 6 by discussing the issues we faced with annotation and some future research directions.


\section{Related Work}
  Over the years, there has been a lot of interest in query segmentation in the search engine space. Few of the initial query segmentation work were based on computing mutual information on the query terms. \citeauthor{risvik2003query} \cite{risvik2003query} segments the query using connexity values that use the mutual information within a segment along with the segment's frequency in the query logs. \citeauthor{huang2010exploring} \cite{huang2010exploring} use segment based Pairwise Mutual Information (PMI) to build a web-scale language model. Among other mutual information based techniques, \citeauthor{jones2006generating} \cite{jones2006generating} find query segments for effective query substitution. There are several other techniques to find effective query segments based on Conditional Random Fields. \citeauthor{kiseleva2010unsupervised}\cite{kiseleva2010unsupervised} user users' click data along with the query to segment the queries in an unsupervised way. \citeauthor{guo2008unified}\cite{guo2008unified} use CRF techniques to segment the queries more as a query refinement task. This task involves spelling correction, stemming etc. as tasks that are solved in parallel to the query segment tasks which result in effective query refinement. Most of the current query segmentation techniques  assume the queries to be corrected from a spell checker and stemmed if required. Using word embeddings, we will show that spelling corrections are handled implicitly while segmenting the queries and the semantic model is immune to common spelling mistakes.
  
  \citeauthor{tan2008unsupervised} \cite{tan2008unsupervised} and \citeauthor{zhang2009query} \cite{zhang2009query} suggest unsupervised methodologies to find quality query segments. \citeauthor{tan2008unsupervised} use the raw query frequencies using the Wikipedia corpus to build a language model via expectation maximization. They then boost the segmentation scores derived form this language model for a query segment if its prominently featured in Wikipedia.
\citeauthor{bergsma2007learning} \cite{bergsma2007learning} treated query segmentation as a supervised learning tasks handcrafting few features like POS tag features, statistical features like query and phrase frequencies in web and query logs, context dependent features focusing on noun phrases. The authors also share 500 queries as a gold data set annotated by 3 human judges. Many subsequent works have used this data set. 500 queries is a rather small dataset for a web search query segmentation task and its hard to prove the generalization power of the model on it. \citeauthor{hagen2011query} \cite{hagen2011query} introduce a 50K gold data set annotated by 10 human judges.  We use this gold data set for benchmarking query segmentation accuracies on web search queries. \citeauthor{hagen2010power} \cite{hagen2010power} introduce a query segmentation method using the raw n-gram frequencies of the segments in the search query logs. They introduce a scoring function which computes the weighted sum of frequencies contained in the n-grams of a query. In their subsequent  
work \cite{hagen2011query} they boost the segmentation scores  for the phrases that appear as Wikipedia titles on similar lines to \citeauthor{tan2008unsupervised} \cite{tan2008unsupervised}. \citeauthor{parikh2013segmentation} \cite{parikh2013segmentation} used the same n-gram based scoring function \cite{hagen2010power} on eBay eCommerce queries. They  also introduce a few evaluation metrics to measure the quality  of the query segments from an eCommerce point of view. 

The scoring function using raw n-gram frequencies mentioned in \citeauthor{hagen2011query} \cite{hagen2011query} is an unsupervised approach which gives higher weight to long segments compared to shorter ones in order to compensate the power law distribution of occurrence frequencies. Though this is true in most cases, it does not work well for variations in word order for the same query. \citeauthor{bergsma2007learning} \cite{bergsma2007learning} show that supervised techniques work well for query segmentation task but it requires handcrafting of features capturing the essence of the underlying data set. The only work to the best of our knowledge which talks about query segmentation in eCommerce setup is from \citeauthor{parikh2013segmentation} \cite{parikh2013segmentation}. They use the naive n-gram scoring function \cite{hagen2010power} on eCommmerce queries.

\section{Motivation}
	Word embeddings techniques such as word2vec \cite{mikolov2013distributed}, Glove \cite{pennington2014glove} and fasttext \cite{bojanowski2016enriching} are used to capture low dimensional space representation of words. These representations help to learn semantics while they optimize for word context with their window-based training style. The output probabilities predict how likely it is to find each vocabulary word nearby the input word.
    
    Our approach uses this context optimization to identify the best possible segment boundaries in the query. Words in one segment n-gram appear more often in queries in the same context compared to words which do not belong to one segment. This helps to optimize for higher similarity for words in potentially same segment versus others in the vocabulary. Levy et al. demonstrate that word2vec implicitly factorizes a word-context PMI matrix \cite{levy2015pmi}. PMI is known to be high for frequently co-occurring terms and low otherwise. This is the primary motivation of our approach to utilize word2vec style algorithms to learn query embeddings for segmentation.

\begin{figure}[ht]
\centering
\includegraphics[width=.45\textwidth]{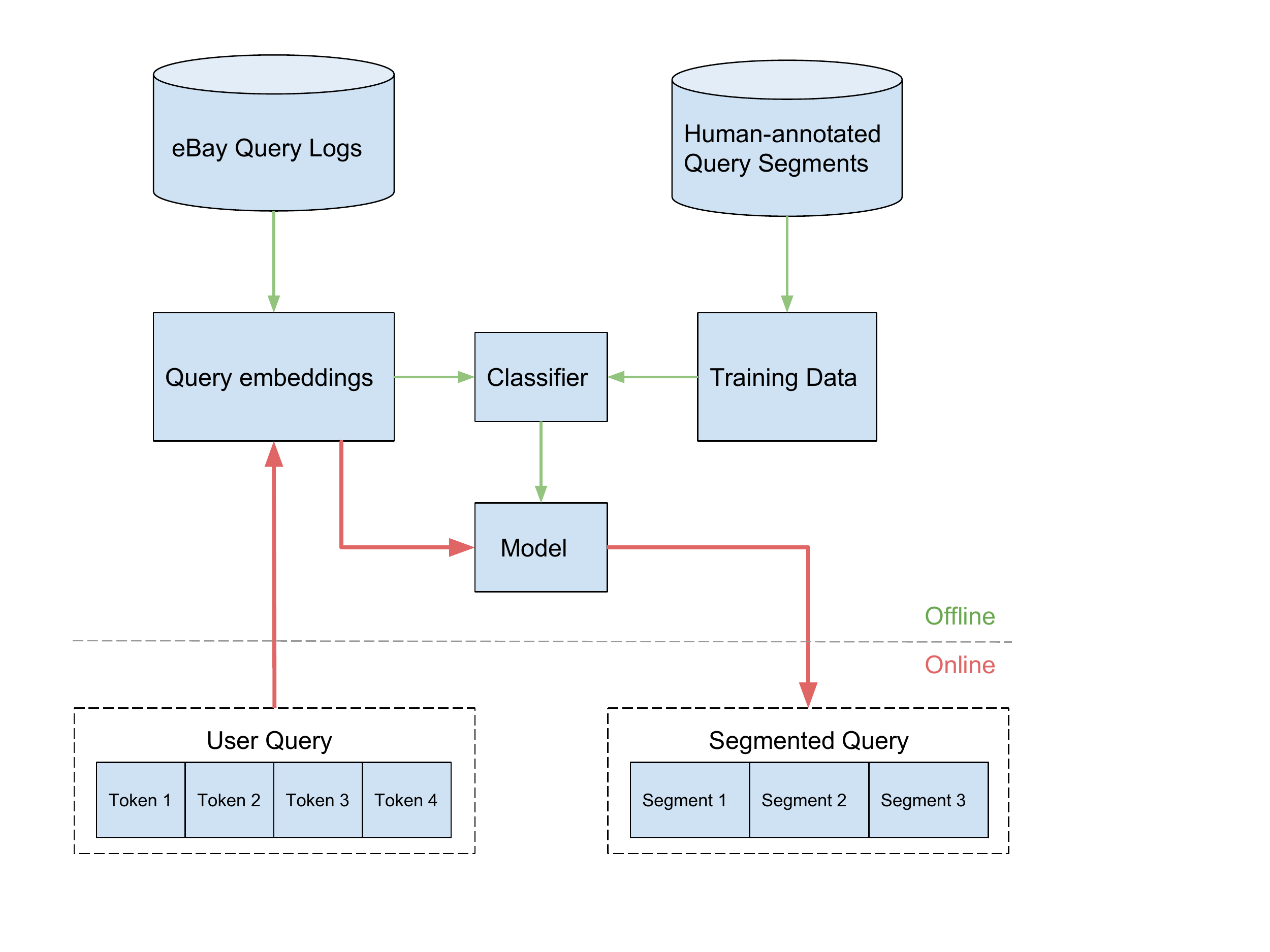}
\vspace{-2mm}
\caption{Query Segmentation architecture}
\vspace{-2mm}
\label{fig:architecture}
\end{figure}

\section{Classifier Architecture}
  In this section we describe query segmentation as a supervised learning task using query embeddings as input features to the classifier. For a query with \textbf{\textit{N}} tokens, we will have \textbf{\textit{N-1}} boundaries between the tokens. We train a binary classifier to make these \textbf{\textit{N-1}} segmentation decisions.  
	
    We demonstrate word embeddings model on two sets of query logs: web search queries and eBay eCommerce queries. As shown in Figure \ref{fig:architecture}, after extracting queries from the query logs, we train the query embeddings. Section 5 will go into details on the embeddings learning parameters.   
    
  Consider a query \textbf{Q = $W_1$ $W_2$ $W_3$ $W_4$ ... $W_N$  } with \textbf{N} tokens. We model the query segmentation task as an independent decision making task to draw a segmentation break between the two words. For every pair of words \textbf{$W_i$ , $W_{i+1}$} in a query we concatenate the 2 vectors (\textbf{\textit{D}} dimension each) representing \textbf{$W_i$} and \textbf{$W_{i+1}$} into a \textbf{\textit{2D}} vector. This concatenated vector is the input to the binary classifier which decides to segment or not between \textbf{$W_i$} and \textbf{$W_{i+1}$}.
  
\section{Experiments}
  
\subsection{Performance Metrics}
  We segment the queries and measure the performance of our model based on two metrics: 
  
\begin{itemize}
\item \textit{Segmentation Accuracy}: Segmentation accuracy is the percent of the correct decisions made - whether to apply segmentation break or not for a combined set of segments in all queries. Its global score for the entire query set. 
\item \textit{Query Accuracy}: Query accuracy is the percent of the correctly segmented queries. This is a stricter metric as the classifier has to achieve all the segments within a query to get credit.
\end{itemize}

\subsection{Experiments on web search query data}
\begin{figure}[ht]
\centering
\includegraphics[width=.45\textwidth]{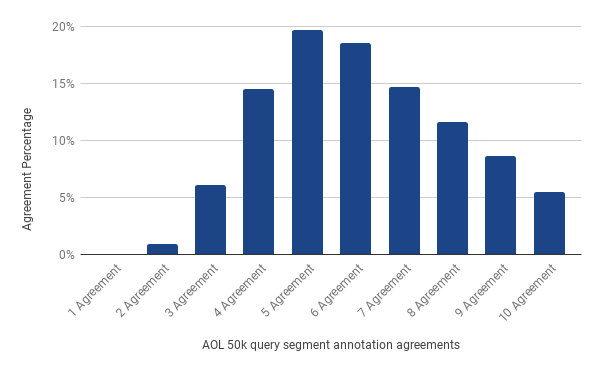}
\vspace{-2mm}
\caption{AOL web search query annotation agreement}
\vspace{-2mm}
\label{fig:aol_annotation_dist}
\end{figure}

\citeauthor{bergsma2007learning} \cite{bergsma2007learning} choose 1500 queries from 36M AOL query set \cite{pass2006picture} and divide the data set into train, test, validation sets (500 each) to evaluate their segmentation techniques. As mentioned in  \citeauthor{hagen2010power} \cite{hagen2010power} the data set is small to perform extensive comparative analysis especially for web search queries which have a wide domain distribution. For a much broader analysis of the segmentation algorithm effectiveness \citeauthor{hagen2011query} \cite{hagen2011query} annotated 50K AOL queries from the same 36M AOL corpus using 10 annotators. We use this dataset to test our approach. Figure \ref{fig:aol_annotation_dist} shows a distribution of annotator agreement for this corpus.

  We do a 60-20-20 train-val-test split on the 50k queries. To train the query embeddings, we remove the test set from the 36M AOL queries. We use gensim \cite{rehurek_lrec} word2vec wrapper to build the model. Since queries are short texts we use a smaller context window (2-3). Our experiments show this smaller context window learns better embeddings for queries especially for the segmentation task. We set the dimension to 300, which leads to a 600 dimensional vector when we concatenate the 2 vectors. This 600 dimensional vector becomes the input feature vector to the classifier. Once the embeddings model is built, we build a classifier on the training set and tune the parameters using validation set. We tried multiple classifiers but  Logistic Regression and XGBoost \cite{chen2016xgboost} performed the best. We found higher lift in the segmentation accuracy when we use GBMs. We report these numbers in Table \ref{tab:word_emb_aol_best_wo_boost}.
%
%

For benchmarking we build naive query segmentation model \cite{hagen2010power} over the same data set. We can see that a simple logistic regression on the embeddings vector gives better result than the raw ngram frequency boosting technique (Table \ref{tab:word_emb_aol_best_wo_boost}). We need to note that the authors propose an improved model with boosting segments using Wikipedia Titles \cite{hagen2011query}, but we dont use it to keep the models agnostic of external data augmentation. We plan to explore that as part of our future work. 
  Google's word2vec \cite{mikolov2013distributed} released pretrained word embedding vectors on Google News. These pretrained embeddings were built on data set which has about 100 Billion words. It has has around 3 million words and phrases, each having 300 dimensional vector. 
  
  We explore the idea of transfer learning to use these pretrained embeddings (from news data corpus) for web query segmentation. We follow a similar process as we did before when we trained a embeddings model from scratch. These pretrained vectors act as features for each word in a query, concatenate them and feed it to the classifier. XGBoost was used as the classifier and we performed a grid search to find the best hyper-parameters for depth of the tree and number of estimators. This process was repeated with pretrained GloVe vectors on common crawl \cite{pennington2014glove} and facebook fasttext \cite{bojanowski2016enriching} pretrained model over Wikipedia corpus with 2.5M word vocabulary. \ref{tab:word_emb_aol_best} shows the experiment results. We got the best performance with Glove web crawl pretrained embeddings. These numbers are reported in Table \ref{tab:word_emb_aol_best}.
  
\begin{table}[!htb]
\captionsetup{size=footnotesize}
\caption{Segmentation and Query accuracies on AOL queries} \label{tab:word_emb_aol_best_wo_boost}
\setlength\tabcolsep{0pt} 
\footnotesize\centering
\smallskip 
\begin{tabular*}{\columnwidth}{@{\extracolsep{\fill}}rcccr}
\toprule
  Method   &  Segmentation Accuracy & Query Accuracy\\
\midrule
Naive n-gram  &  0.677 & 0.351 \\
Logistic Regression  & 0.731  & 0.418 \\
\midrule
\bottomrule
\end{tabular*}
\end{table}

\begin{table}[!htb]
\captionsetup{size=footnotesize}
\caption{Best accuracies using pretrained word embeddings using XGBoost classifier (on AOL queries)} \label{tab:word_emb_aol_best}
\setlength\tabcolsep{0pt} 
\footnotesize\centering
\smallskip 
\begin{tabular*}{\columnwidth}{@{\extracolsep{\fill}}rcccr}
\toprule
 Word Vectors  &  Segmentation Accuracy & Query Accuracy\\
\midrule
GloVe web crawl  & \textbf{0.811} & \textbf{0.552} \\
GloVe web crawl {average}   &  0.781 & 0.497  \\  
Google News word2vec  & 0.797  & 0.523 \\
Facebook Fasttext wikipedia  &  0.806 & 0.548  \\
word2vec model on AOL corpus  & 0.804 & 0.54    \\


 \midrule
\bottomrule
\end{tabular*}
\end{table}

%
%

\subsection{Experiments on eCommerce query data}
\begin{figure}[ht]
\centering
\includegraphics[width=.45\textwidth]{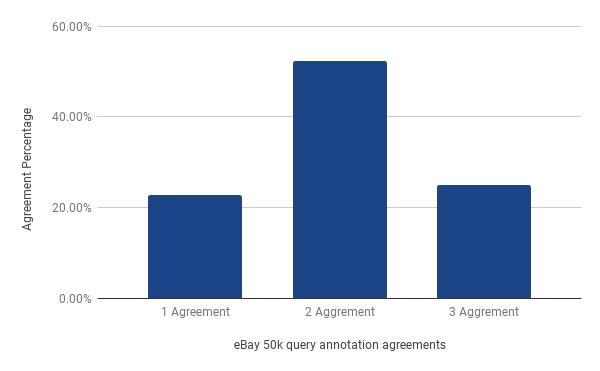}
\vspace{-2mm}
\caption{eBay query annotation agreement}
\vspace{-2mm}
\label{fig:ebay_annotation_dist}
\end{figure}

We annotated a 50000 query set sampled across top 15k eBay categories over a period of 3 months. Each query received 3 annotations each with meaningful segments keeping  search relevance in mind. For precision the annotators were asked to verify the segmented query on eBay's search engine to ensure quality of results improve post segmentation. Figure \ref{fig:ebay_annotation_dist} shows the annotator agreement distribution. We observed, 2 or more agree on around 77\% of queries whereas all 3 annotators are in agreement around 25\% times. As others have found out in the past experiments, query segmentation has lot of ambiguity around it which reflects in the difference of opinion in the annotations. A low percent on all 3 in agreement highlights the fact that segmentation tasks can be very subjective.
%
%
%
%

For training we use the annotations where at least 2 out of 3 annotators are in agreement.

We divide this query set into 80-20\% train and test split. The 80\% training data is further split into 80-20\% train and validation set.

As shown in Figure \ref{fig:architecture} the query embeddings were learned using fasttext followed by training an XGBoost \cite{chen2016xgboost} classifier for predicting the segmentation breaks. A grid search for hyper parameter tuning got us the best segmentation accuracy at 500 estimators and with depth 4 on the XGBoost model with skip-gram and 800 estimators with depth 6 for cbow embeddings. We experimented with both cbow and skip-gram architectures for learning the query embeddings on eBay queries. The best numbers for both query accuracy and segmentation accuracy were using cbow architecture. Figure \ref{tab:ebay_accuracies} summarizes the accuracies for the naive ngram model and our approach. The accuracy we obtain on the eCommerce data using our technique clearly beats the naive ngram based  segmentation model. These numbers are also comparable with the embeddings experiment which was ran on AOL queries. 

\begin{table}[!htb]
\captionsetup{size=footnotesize}
\caption{Segmentation and Query accuracies on eBay queries} \label{tab:ebay_accuracies}
\setlength\tabcolsep{0pt} 
\footnotesize\centering
\smallskip 
\begin{tabular*}{\columnwidth}{@{\extracolsep{\fill}}rcccr}
\toprule
Method   &  Segmentation Accuracy & Query Accuracy\\
\midrule
Naive n-gram  &  0.713 & 0.578 \\
Fasttext skip-gram + XGBoost  & 0.796  & 0.677 \\
Fasttext cbow + XGBoost & \textbf{0.799} & \textbf{0.683} \\
\midrule
\bottomrule
\end{tabular*}
\end{table}

%
%

\section{Discussion and Future Work}
In our study, we found eCommerce queries are harder for a crowd sourcing annotation task because of the lack of product and domain knowledge across all shopping categories which can be critical for understanding the query intent before annotating. For example consider the query \textit{"infant lebron"}. This query does not seem obvious for the annotators to identify the query segment. Because, they need to know that \textit{lebron} is a brand of athletic shoes and \textit{infant} is actually the size description. We noticed a few annotators contesting such queries asking for more clarity.

For queries like \textit{"toms men s size 12"}, the user is searching for \textit{men's} but some annotators misinterpret it as \textit{(size) S} and decide to have a segmentation break between \textit{men} and \textit{s} but the model predicts such scenarios right by keeping \textit{men} and \textit{s} together as a single segment. There is lot of ambiguity what is a head object of a query and what is not. For example \textit{"food processor"} or \textit{"coffee grinder"} clearly belong together as one segment, but it is debatable if \textit{"motorcycle vest"} or \textit{"golf pants"} necessarily belong in one segment. One can argue \textit{motorcycle} is just a constraint applied to buy a particular type of  \textit{vest} and \textit{"motorcycle vest"} together does not necessarily represent the head shopping object. Aspects or modifiers of a query are tricky and the ambiguity leads to a drop in the classifier accuracy. 

As pointed out by \citeauthor{hagen2011query} \cite{hagen2011query}, the quality and metrics used for segmentation task heavily depend on the downstream task. A deeper analysis of our downstream tasks is an area we intend to explore in our future work. There has been some work on boosting the segmentation score with Wikipedia entities and Wikipedia titles, we plan to explore using structured data from eCommerce dictionary like brands, colors, styles and other aspects to further augment the training data for the embeddings. We also observed the n-gram based approaches are very sensitive to word order in the queries. For example \textit{"yeezy boost 350"} is a product line offering by \textit{adidas} and the model keeps it together in one segment, but for variation in the order like \textit{"boost 350 yeezy"} (which is one of the high frequency query we observed) the naive n-gram model fails to keep it together and wrongly segments it as \textit{boost 350 \textbf{|} yeezy } but our approach generalizes pretty well on word order. The context based cost function in all the query embedding approaches we took would explain that behavior.

\section{Acknowledgements}
  We would like to thank Matthias Hagen, Martin Potthast, Benno Stein and Christof Brautigam, authors of \citeauthor{hagen2011query} \cite{hagen2011query} for sharing the 50K annotated web search query  data set.

\bibliographystyle{ACM-Reference-Format}
\bibliography{sigproc} 

\end{document}